\documentclass[10pt, conference, A4]{IEEEtran}
\IEEEoverridecommandlockouts

\usepackage[utf8]{inputenc}
\usepackage{amsmath}
\usepackage{amssymb}
\usepackage{amsfonts}
\usepackage{bm}
\usepackage{graphicx} 
\usepackage{adjustbox}
\usepackage{wrapfig}
\usepackage{array}
\usepackage{multirow}
\usepackage{booktabs}
\usepackage{threeparttable}
\usepackage{tablefootnote}
\usepackage{balance}
\usepackage{textcomp}
\usepackage{url}
\usepackage{xurl}
\usepackage{microtype}

\usepackage{tikz}
\usepackage{tikzsymbols}
\usetikzlibrary{automata, positioning, shapes, fit, calc, shadows, shapes.geometric, arrows, arrows.meta, patterns, decorations.markings, decorations.pathreplacing, pgfplots.groupplots}
\usepackage{pgfplots}
\pgfplotsset{compat=1.18}
\usepackage{circuitikz}

\usepackage[linesnumbered,ruled,vlined]{algorithm2e}
\usepackage{algorithmic}
\algsetup{linenosize=\small}

\usepackage{enumitem}
\usepackage{titlesec}
\usepackage[skip=-1pt, font=small]{caption}
\usepackage{subcaption}
\usepackage{floatrow}
\newfloatcommand{capbtabbox}{table}[][\FBwidth]

\usepackage[hidelinks]{hyperref}
\hypersetup{
    colorlinks=true,
    linkcolor=black,
    citecolor=blue,
    urlcolor=blue
}

\usepackage{xcolor}
\usepackage{color}
\usepackage{pifont}
\usepackage{soul}
\usepackage{microtype}
\usepackage{comment}
\usepackage{stfloats}
\usepackage{orcidlink}

\usetikzlibrary{svg.path}

\definecolor{orcidlogocol}{HTML}{A6CE39}
\tikzset{
    orcidlogo/.pic={
        \fill[orcidlogocol] svg{M256,128c0,70.7-57.3,128-128,128C57.3,256,0,198.7,0,128C0,57.3,57.3,0,128,0C198.7,0,256,57.3,256,128z};
        \fill[white] svg{M86.3,186.2H70.9V79.1h15.4v48.4V186.2z}
        svg{M108.9,79.1h41.6c39.6,0,57,28.3,57,53.6c0,27.5-21.5,53.6-56.8,53.6h-41.8V79.1z M124.3,172.4h24.5c34.9,0,42.9-26.5,42.9-39.7c0-21.5-13.7-39.7-43.7-39.7h-23.7V172.4z}
        svg{M88.7,56.8c0,5.5-4.5,10.1-10.1,10.1c-5.6,0-10.1-4.6-10.1-10.1c0-5.6,4.5-10.1,10.1-10.1C84.2,46.7,88.7,51.3,88.7,56.8z};
    }
}

\newcommand\orcidicon[1]{\href{https://orcid.org/#1}{\mbox{\scalerel*{
                \begin{tikzpicture}[yscale=-1,transform shape]
                \pic{orcidlogo};
                \end{tikzpicture}
            }{|}}}}

\def\BibTeX{{\rm B\kern-.05em{\sc i\kern-.025em b}\kern-.08em
    T\kern-.1667em\lower.7ex\hbox{E}\kern-.125emX}}

\begin{document}
\bstctlcite{IEEEexample:BSTcontrol}

\title{Security Analysis of Universal Circuits as a Mechanism for Hardware Obfuscation}

\author{
\IEEEauthorblockN{Zain~Ul~Abideen\textsuperscript{\orcidlink{0000-0002-8865-9402}} $^{\dagger}$, 
Deepali~Garg\textsuperscript{\orcidlink{0009-0003-3304-9599}} $^{\ddagger}$, Lawrence~Pileggi\textsuperscript{\orcidlink{0000-0002-8605-8240} $^{\ddagger}$}, and 
Samuel~Pagliarini\textsuperscript{\orcidlink{0000-0002-5294-0606}} $^{\ddagger}$} 
\IEEEauthorblockA{
$^{\dagger}$ Department of Electrical and Computer Engineering, University of Idaho, ID, USA\\
$^{\ddagger}$ Department of Electrical and Computer Engineering, Carnegie Mellon University, PA, USA\\
zabideen@uidaho.edu, pagliarini@cmu.edu, \{deepalig, pileggi\}@andrew.cmu.edu}
\vspace{-1.0cm}
}

\maketitle

\begin{abstract}
Universal Circuits (UCs) offer a promising approach to hardware Intellectual Property (IP) obfuscation, leveraging cryptographic principles to hide both structure and function in a programmable logic fabric. Their adaptability makes them especially suitable for the globalized Integrated Circuit (IC) supply chain, where security against threats like reverse engineering is crucial. Despite the potential, UC security remains largely unexplored. This work evaluates UC security against state-of-the-art oracle-guided (OG) and oracle-less (OL) attacks. Results show near-random success rates (\(\approx 50\%\)) for OG attacks whereas OL attacks display minimal structural leakage. Collectively, these findings confirm the feasibility of UCs for IP protection.
\end{abstract}


\vspace{-2pt}
\section{Introduction} \label{sec:intro}
\vspace{-1.5pt}
As Integrated Circuit (IC) fabrication technology advances, smaller technology nodes enable ever-increasing performance at reduced energy costs. However, leading edge silicon is only available through a globalized, distributed IC supply chain. Several potential threats emerge from this scenario~\cite{alkabani2007, roy2008, ramesh2010}. To counter these, the hardware security community has proposed several obfuscation techniques, most notably Logic Locking (LL) and Reconfigurable-based obfuscation (\textsc{REbo})~\cite{abideen2024}. These countermeasures often rely on informal security assumptions that do not hold against certain adversarial models, leading to a never-ending cycle of attacks and defenses. The concept of \textsc{REbo} has gained popularity in recent years as a superior alternative to LL, but there are still concerns as pointed out in recent attacks~\cite{zhaokun2023, rezaei2022, Abideen2025b}.

In this context, a novel obfuscation technique that leverages Universal Circuits (UCs)~\cite{valiant1976, elisaweta2022} has been recently introduced. UCs, due to their inherent flexibility and cryptographic nature, provide a promising solution: UCs hide both the topology and functionality of the circuit being obfuscated. 
UCs can \emph{emulate} any Boolean function or circuit while not requiring any tailoring for that specific circuit. Thus, when using a UC as a locked (obfuscated) circuit, no information is leaked to the end-user as to which specific circuit is being executed. This adaptability is realized through programming bits that configure the gates and the interconnect within the UC, enabling it to assume the behavior of the intended circuit~\cite{valiant1976}. The use of UCs as a means of obfuscation, termed \textbf{UC-based obfuscation} (\textsc{UCbo})~\cite{elisaweta2022}, remains largely understudied. 

The claim that UC implementation effectively ensures the security of IP is substantiated by formal proofs and contemporary research~\cite{elisaweta2022, crescenzo2020}. In general, theoretical advancements in UC constructions have progressed through the use of 2- to 3-input universal gates, including the timely LUT implementation described in~\cite{disser2023}. There are also methodologies for the implementation of UC, including the Field-Programmable Gate Array-based methodology described in~\cite{elisaweta2022}. 

However, questions remain regarding the resilience of \textsc{UCbo} against realistic adversaries and diverse attacks. Presently, there is a lack of detailed security analysis for \textsc{UCbo} in the literature: no work addresses state-of-the-art attacks applicable to LL, \textsc{REbo}, and other obfuscation techniques \cite{subramanyan2015, alaql2021, abideen2024} and whether they can be used for deobfuscating \textsc{UCbo}. This paper is the \textbf{first to perform a security analysis} of \textsc{UCbo}, demonstrating high attack complexity, low key recovery rates, and practical viability.




\vspace{-2pt}
\section{UC-Based Obfuscation} \label{sec:concept}
\vspace{-1.5pt}
To achieve a provable countermeasure against security threats, we consider a cryptographically inspired hardware obfuscation method based on UCs. The inputs required for the \textsc{UCbo} method in~\cite{elisaweta2022} comprise the original circuit \(C\), which contains \(x\) inputs, \(v\) outputs, a size of \(s\), and a security parameter \(\lambda\). The locking process utilizes \(\lambda\) and the original circuit \(C\) to yield a locked circuit denoted by \((p, UC_{x,v}^s)\). 
In this context, the bitstream \( p \) serves as a key that encodes the specific behavior of the circuit \( C \). When \( p \) is supplied to the locked circuit, the original functionality is faithfully reproduced. That is, for any input \( \vec{x} \in \{0,1\}^x \), it holds that
\[
UC_{x,v}^s(p, \vec{x}) = C(\vec{x}).
\]
The UC structure \( UC_{x,v}^s \) ensures that the obfuscated circuit reveals no non-trivial information about the original design. The amount of information that \emph{leaks} from the UC is formalized as computational indistinguishability in~\cite{thesis}. This ensures that the obfuscated representation provides no advantage over black-box interaction with the original circuit. For readers familiar with LL literature, this implies structural attacks are bound to fail with a very high probability.



The obfuscation process involves embedding \( C \) into a programmable UC fabric using an edge-universal graph (EUG), which serves as the structural backbone to accommodate arbitrary topologies up to size \( s \), as shown in Algorithm~\ref{algo:ucbo}.
\begin{algorithm}[!htb]
\DontPrintSemicolon
\small

\KwIn{Circuit $C$ with $n$ inputs, $m$ outputs, and $s$ gates}
\KwOut{Obfuscated circuit $C^{\text{UC}}$, configuration string $p$}

Parse $C$ into a directed acyclic graph (DAG) $G(V,E)$;

Construct an edge-universal graph $G^*$ of size $s$;

\ForEach{node $v \in G^*$}{
  \eIf{$v$ corresponds to a logic node}{
    Replace with 2-input universal gate\;
  }{
    Replace with a programmable switchbox or buffer\;
  }
}
Add programmable interconnects for all edges\;
$p \leftarrow \texttt{Program}(C)$ such that $C^{\text{UC}}(p, x) = C(x)$\;
\Return{$C^{\text{UC}}, p$}
\caption{UC-Based Obfuscation of Circuit $C$}
\label{algo:ucbo}
\end{algorithm}

This algorithm ensures that the functional behavior of the obfuscated circuit remains identical to that of the original, while its physical and structural representation becomes independent of the original logic. In other words, many possible circuits can map to the same universal structure, and without access to $p$, the adversary cannot distinguish which function is intended. 


\vspace{-2pt}
\subsection{Design Obfuscation Flow} \label{sec:methodology}
\vspace{-1.5pt}
This section explains how to map UCs to Application-specific Integrated Circuits (ASICs) following the approach described in~\cite{elisaweta2022}. The most straightforward approach to implement UCs would be to feed universal gates and switches directly to ASIC synthesis as the components of a library. However, this is not feasible as synthesis tools are not security aware. To overcome this limitation, a framework that automatically maps UC constructions using a custom reconfigurable architecture is appropriate -- one that borrows LUT-based concepts from FPGA flows in order to build an ASIC flow.

\tikzstyle{title} = [fill=white,font=\fontsize{8}{8}\color{black!50}\ttfamily]
\tikzstyle{block} = [draw, fill=white, rectangle, align=center, minimum height=1em, minimum width=5em, font=\fontsize{8}{7}\color{black}\ttfamily]
\tikzstyle{smallblock} = [draw, fill=white, rectangle, align=center, minimum height=1em, minimum width=5em, font=\fontsize{8}{7}\color{black}\ttfamily]
\tikzstyle{specialblock} = [draw, fill=white, rectangle, align=center, minimum height=1em, minimum width=5em, font=\fontsize{8}{7}\color{black}\ttfamily]

\tikzstyle{logo} = [draw, fill=white, rectangle, 
    minimum height=3em, minimum width=3em]
\tikzstyle{stacked} = [double copy shadow, shadow xshift=2pt, shadow yshift=-2pt, every edge/.append style = {draw, semithick, -Stealth}]

\begin{figure}[htb]
\begin{tikzpicture}[auto, node distance=2cm,>=latex']

\node [specialblock, name=rtlsso, yshift=-15pt] {Parser\\ \textcolor{gray}{LUT Format}};
\node [specialblock, name=yes, below of = rtlsso, xshift=0pt, yshift=28pt] {\textsc{Optimized HW}\\ \textcolor{gray}{LUTs, EUG}};
\node [specialblock, name=ucompile, below of = yes, xshift=0pt, yshift=28pt] {VPR 8\\ \textcolor{gray}{Reconfigurable Fabric}};
\node [block, thick, name=circuit, left of = rtlsso, xshift = -15pt, yshift=0pt] {Circuit \\ \textcolor{gray}{Verilog}};
\node [specialblock, name=outputAsset, below of = ucompile, yshift=28pt,xshift=0pt] {Logic Synthesis\\ \textcolor{gray}{Commercial Technology} };
\node [block, name=outputTB, below of = outputAsset, yshift=28pt,xshift=0pt] {Physical Synthesis\\ \textcolor{gray}{Lib + Lef} };

\node [block, name=outputMetric, below of = outputAsset, yshift=0pt, xshift=-35pt] {Layout\\ \textcolor{gray}{signoff} };

\node [block, name=outputMetric1, below of = outputAsset, yshift=0pt, xshift=35pt] {Results\\ \textcolor{gray}{PPA} };


\node[anchor=center, draw=gray, thick, inner sep = 0, circle, minimum size=2.6em]
    (icon_rtlsso) at ($(rtlsso.east) + (2.45, 0)$) {\includegraphics[width=15pt]{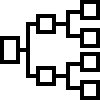}};
\draw[thick, ->] (rtlsso.east) -- (icon_rtlsso.west);

\node[anchor=center, draw=gray, thick, inner sep = 0, circle, minimum size=2.6em]
    (icon_yes) at ($(yes.east) + (2.3, 0)$) {\includegraphics[width=18pt]{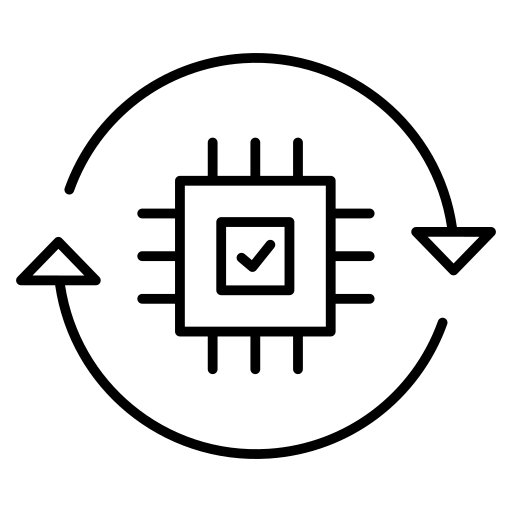}};
\draw[thick, ->] (yes.east) -- (icon_yes.west);

\node[anchor=center, draw=gray, thick, inner sep = 0, circle, minimum size=2.6em]
    (icon_vpr) at ($(ucompile.east) + (1.55, 0)$) {\includegraphics[width=18pt]{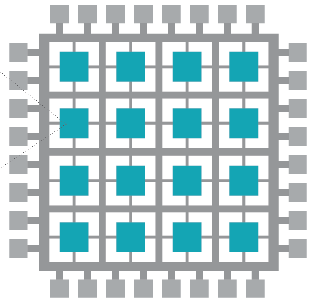}};
\draw[thick, ->] (ucompile.east) -- (icon_vpr.west);

\node[anchor=center, draw=gray, thick, inner sep = 0, circle, minimum size=2.6em]
    (icon_asset) at ($(outputAsset.east) + (1.5, 0)$) {\includegraphics[width=18pt]{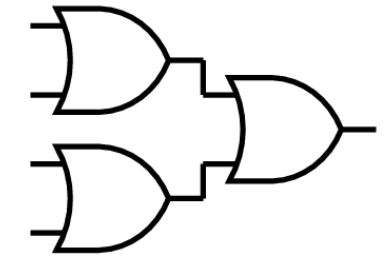}};
\draw[thick, ->] (outputAsset.east) -- (icon_asset.west);

\node[anchor=center, draw=gray, thick, inner sep = 0, circle, minimum size=2.6em]
    (icon_tb) at ($(outputTB.east) + (1.75, 0)$) {\includegraphics[width=18pt]{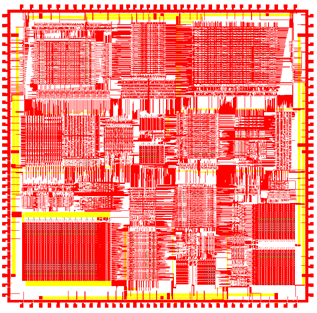}};
\draw[thick, ->] (outputTB.east) -- (icon_tb.west);

\draw[thick, ->, black] (circuit)  edge ($(rtlsso.west) + (0.0, 0)$);
\draw[thick, ->, black] (rtlsso)  edge  ($(yes.north) + (0.0, 0)$);
\draw[thick, ->, black] (yes)  edge  ($(ucompile.north) + (0.0, 0)$);
\draw[thick, ->, black] (outputAsset)  edge  ($(outputTB.north) + (0.0, 0)$);

\node[font=\fontsize{8}{8}\color{black}\ttfamily, right of = rtlsso, yshift=-15pt,xshift=-48pt] {EUG (Poles)};

\node[font=\fontsize{8}{8}\color{black}\ttfamily, right of = yes, yshift=-20pt,xshift=-14pt] {};
 
\draw[->, black] (outputTB) edge (outputMetric);

\draw[thick, ->, black] (ucompile)  edge (outputAsset);

\draw[->, black] (outputTB) edge (outputMetric1);

\end{tikzpicture}
\caption{Flow for the ASIC implementation.}
\label{fig:concept}
\end{figure}
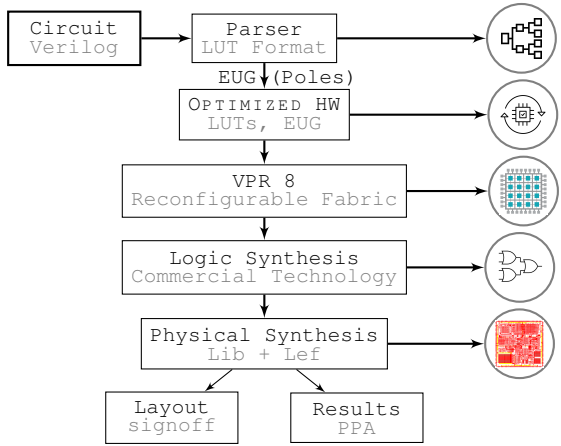

Figure~\ref{fig:concept} provides an overview of the LUT-based obfuscation flow, which begins with a Verilog parser that reads the targeted design and converts it into a LUT-based combinational circuit format. UC mappings are designed to embed any circuit with a given fan-in and fan-out degree $k$. In order to achieve this, the input netlist is pre-processed using synthesis tools (e.g., Yosys). This step converts the circuit to comprise only of LUTs up to the given input size. Next, circuit nodes with a fan-out greater than $k$ have buffers added between them to ensure each gate has a fan-out of no more than $k$.

The complexity of the circuit mapping is determined by the number of poles in the EUG and the LUT mapping helps reduce these poles to minimal. Next, in the optimized HW stage, the number of LUTs(n) along with inputs(s) are used to determine the EUG configuration for a given fan-in degree $k$. This ${EUG}_k(n + s)$ can be mapped onto a reconfigurable fabric which requires only $\mathcal{O}((n + s)^2)$ CLBs. 

Following this, VPR is utilized to determine programming bits for the fabric generated by the optimized HW step. These bits define both the logic functions within the CLBs as well as the routing paths between them. Next, logic synthesis for this reconfigurable fabric is performed using a commercial standard cell library. Finally, physical synthesis is carried out to generate the final layout of the circuit. This includes the placement of standard cells and the detailed routing of signals while considering congestion, timing, etc. 

\vspace{-2pt}
\section{Results} \label{sec:results}
\vspace{-1.5pt}
To evaluate the security of \textsc{UCbo} in a reproducible way, we employ the \emph{ISCAS'89} benchmark suite and an additional CMP design, which is a large synthetic design that combines multiple \emph{ISCAS'89} instances. The adopted benchmarks grow in size considerably when converted to a UC, thus making for representative attack scenarios (the largest design we consider has 200K+ gates). After obfuscating the designs, we converted the circuits to their unrolled combinational equivalents, as many of the existing oracle-guided (OG) and oracle-less (OL) attacks do not support sequentiality. Table~\ref{tab:synthesis} presents the logic synthesis (Cadence Genus) results using a 65nm CMOS commercial technology for both the original sequential circuits and their unrolled obfuscated counterparts. The second part of Table~\ref{tab:synthesis} lists the results for the 10-times unrolled obfuscated circuits that were generated using the flow detailed in Fig.~\ref{fig:concept}.


\begingroup
\setlength{\tabcolsep}{2.0pt} 
\renewcommand{\arraystretch}{1.1} 
\begin{table} [htb]
\footnotesize \centering
\caption{Logic synthesis results for the original sequential circuits and unrolled combinational circuits from the ISCAS’89 suite.}
\vspace{-3pt}
\label{tab:synthesis}
\begin{tabular}{|p{0.9cm}|p{0.9cm}|p{0.9cm}|p{0.9cm}|p{0.9cm}|p{0.9cm}|p{0.9cm}|} \hline
\multirow{2}{*}{\textbf{Cir.}} & \multicolumn{3}{c|}{\textbf{Original Circuits}} & \multicolumn{3}{c|}{\textbf{Unrolled 10X + UC}} \\ \cline{2-7}
& \textbf{Area ($\mu m^2$)} & \textbf{Seq. cells} & \textbf{Comb. cells} & \textbf{Key Size (K)} & \textbf{Area ($\mu m^2$)} & \textbf{Comb. cells} \\ \hline
s27 & 45.7 & 3 & 5 & 244 & 2071.4 & 389 \\ \hline
s298 & 266.4 & 14 & 29 & 4529 & 42106.7 & 8321 \\ \hline
s344 &  322.9  & 15 & 39 & 6960 & 64918.4 & 12757 \\ \hline
s382 &  384.9 & 21 & 36 & 6549 & 61077.9 & 12032 \\ \hline
s386 &  263.4 & 6 & 51 & 6455 & 59961.1 & 11852 \\ \hline
s420 & 336.7 & 16 & 31 & 7023 & 65379.6 & 12845 \\ \hline
s444 &  391.3 & 21 & 40 & 6764 & 63373.9 & 12581 \\ \hline
s526 &  440.3 & 21 & 56 & 7831 & 73434.8 & 14528 \\
\hline
CMP & 4186.5 & 201 & 486 & 79966 & 927558 & 212499 \\
\hline
\end{tabular}
\end{table}
\endgroup


\vspace{-3pt}
\subsection{Attack-Trace Metrics} \label{subsec:security_metrics}
\vspace{-1.5pt}
Before proceeding with the detailed security analysis, we establish a set of quantitative metrics to evaluate \textsc{UCbo} security against OG attacks. These metrics are proposed to highlight different aspects of security hardness and computational complexity introduced by obfuscation. Specifically, we consider four metrics: \textit{Normalized Clauses-to-Variables Ratio (CTVR)}, \textit{Clause Growth Factor (CGF)}, \textit{Key-to-Complexity Ratio (KCR)}, and \textit{Normalized Delta Redundancy Ratio ($\Delta$RR)}. Collectively, these metrics provide a comprehensive view of Boolean satisfiability (SAT) solver saturation, structural inflation, key input stress, and the relative stability of structural redundancy. While many LL strategies to thwart SAT have been proposed, it remains an important attack -- any obfuscation strategy that does not clear the bar of a SAT attack is inherently weak. Furthermore, all metrics are computed from the CNF characteristics generated by SAT solvers. Importantly, we interpret them together with attack outcomes, not as standalone security guarantees.

\textbf{CTVR \& Normalized CTVR}: CTVR is a ratio that provides insights into the Boolean constraint density embedded in the CNF formula used during the SAT-solving process. A higher CTVR reflects a tighter encoding with fewer degrees of freedom, thereby increasing the search effort required by the SAT solver. To allow for comparative analysis across designs and attack configurations, we normalize CTVR using min-max normalization. This normalization bounds values between 0 and 1 for fair visualization across circuits of different sizes. The CTVR and the Normalized CTVR are given below:
\begin{align}
\text{CTVR}_{\text{attack}}^{(i)} = \frac{C_{\text{attack}}^{(i)}}{V_{\text{attack}}^{(i)}},  && \text{CTVR}^{\text{norm}} = \frac{\text{CTVR}_{\text{attack}}^{(i)} - \min}{\max - \min}
\end{align}
\textbf{CGF \& KCR}: CGF quantifies the increase in clause count of the obfuscated circuit with respect to its unprotected version. A CGF of 1 implies no increase in clause complexity, while higher values reflect the inflation factor due to obfuscation. This metric captures the structural bloat that the attack must handle during CNF processing. CGF alone does not imply security, since clause growth can also come from redundant or highly dependent constraints. KCR is a metric that reflects the ``stress'' placed on the solver in terms of key space relative to problem complexity. It measures the number of inserted key bits against the total number of variables in the CNF formula. The CGF and the KCR are computed as:
\begin{align}
\text{CGF}_{\text{attack}}^{(i)} = \frac{C_{\text{obf}}^{(i)}}{C_{\text{base}}^{(i)}}, 
 && \text{KCR}_{\text{attack}}^{(i)} = \frac{K^{(i)}}{V_{\text{attack}}^{(i)}}
\end{align}
KCR shows the size of the configuration key compared to the SAT variable set for each attack instance. We use this as a stand-in for key pressure and consider it together with runtime and recovery results.

\textbf{$\Delta$RR \& Normalized $\Delta$RR}: $\Delta$RR offers insight into structural redundancy by comparing the number of clauses to variables. We observe that its absolute value often exhibits similar trends across attacks. To better highlight differences between attacks, $\Delta$RR shows how much the RR value changes across different attacks for each circuit. For each circuit, we normalize the spread of RR across all attacks. $\Delta$RR and Normalized $\Delta$RR for a given circuit $i$ are computed as:
\begin{align}
\text{RR}_{\text{attack}}^{(i)} = \left| \frac{C_{\text{attack}}^{(i)} - V_{\text{attack}}^{(i)}}{V_{\text{attack}}^{(i)}} \right| && \Delta \text{RR}_{j}^{\text{norm}} = \frac{\text{RR}_{j} - \min_j}{\max_j - \min_j}
\end{align}
where $j$ indexes the attack type, normalizing for each circuit\footnote{This reveals the circuit's sensitivity to different attack strategies and identifies designs that maintain structural resistance to adversarial attacks.}.

\subsection{Security Analysis} \label{subsec:security_analysis}

Table~\ref{tab:attack_res} summarizes the results of various OG attacks. We have conducted SAT, AppSAT, ATPG-guided SAT, Satisfiability Modulo Theories (SMT)-guided attacks, double differential input pattern (D-DIP), and cyclic SAT (Icy) attacks to provide a comprehensive analysis. The results indicate the complexity of the obfuscated design in terms of variables and clauses. All attacks were executed on an Intel Xeon(R) Platinum 8356H CPU running at 3.90 GHz, with a 48-hour timeout. 

\begingroup
\setlength{\tabcolsep}{2.0pt} 
\renewcommand{\arraystretch}{1.1} 
\begin{table*}[htb]
\footnotesize \centering
\caption{Attack results for SAT~\cite{subramanyan2015}, AppSAT~\cite{shamsi2017}, ATPG-guided SAT~\cite{rajendran2012}, SMT Attack~\cite{zamiri2018}, D-DIP~\cite{shen2017}, Icy~\cite{chian2020} and BBO~\cite{shamsi2019b} attacks on the ISCAS'89 benchmarks circuits are presented. Only s27 was successfully broken.}.
\label{tab:attack_res}
\begin{tabular}{|p{1.7cm}|p{0.8cm}|p{0.8cm}|p{0.8cm}|p{0.8cm}|p{0.8cm}|p{0.8cm}|p{0.8cm}|p{0.8cm}|p{0.8cm}|p{0.8cm}|p{0.9cm}|p{0.8cm}|p{0.6cm}|p{1.58cm}|p{1.58cm}|} 
\hline
\multirow{2}{*}{\textbf{Cir. (I/O)}} & \multicolumn{6}{c|}{\textbf{SAT Variables}} & \multicolumn{6}{c|}{\textbf{SAT Clauses}} & \multicolumn{1}{c|}{\textbf{Time (h)}} & \multicolumn{2}{c|}{\textbf{Bits (Found/Total) - \%}} \\ \cline{2-16}
& \textbf{SAT} & \textbf{APP} & \textbf{ATPG} & \textbf{SMT} & \textbf{D-DIP} & \textbf{Icy} & \textbf{SAT} & \textbf{APP} & \textbf{ATPG} & \textbf{SMT} & \textbf{D-DIP} & \textbf{Icy} & \textbf{Med/Avg} & \textbf{Icy} & \textbf{BBO} \\ \hline
s27 (7/4) & 71217 & 192688 & 71217 & 4289 & 84554 & 56475 & 48008 & 130458 & 48008 & 9977 & 25662 & 161134 & 0.03/0.04 & 131/244-53 & 120/244-49 \\ \hline
s298 (17/20) & 237602 & 161448 & 237596 & 87875 & 184734 & 174067 & 180208 & 119491 & 180192 & 205845 & 456895 & 478142 & TO/TO & 2166/4530-52 & 2250/4530-49 \\ \hline
s344 (24/26) & 309780 & 250629 & 309780 & 135890 & 285599 & 185272 & 233427 & 185933 & 233427 & 318689 & 706853 & 495208 & TO/TO & 3588/6960-51 & 3502/6960-50 \\ \hline
s382 (24/27) & 291071 & 235498 & 291066 & 127681 & 497532 & 254365 & 219209 & 175242 & 219209 & 299402 & 1262390 & 697383 & TO/TO & 3174/6549-51 & 3257/6549-49 \\ \hline
s386 (13/13) & 285315 & 230845 & 285315 & 125478 & 263815 & 248936 & 218572 & 174807 & 218572 & 293980 & 652450 & 683509 & TO/TO & 3130/6455-51 & 3224/6455-49 \\ \hline
s420 (37/17) & 311674 & 252166 & 311674 & 136709 & 532676 & 186573 & 230801 & 182594 & 230801 & 320540 & 1351470 & 498375 & TO/TO & 3720/7023-52 & 3521/7023-50 \\ \hline
s444 (24/27) & 301628 & 244030 & 301628 & 132631 & 516900 & 262693 & 226365 & 180231 & 226365 & 310800 & 1310160 & 720207 & TO/TO & 3503/6765-51 & 3284/6765-48 \\ \hline
s526 (24/27) & 349980 & 283138 & 349980 & 153473 & 322504 & 304815 & 265466 & 211568 & 265466 & 359992 & 798164 & 836371 & TO/TO & 4085/7831-52 & 3932/7831-50 \\ \hline
\textbf{CMP(191/187)} &
2.81M & 2.27M & 2.81M & 1.63M & 4.49M & 2.83M &
2.13M & 1.71M & 2.13M & 3.74M & 11.30M & 7.64M &
TO/TO & 39k/79k-48.8 & 40k/79k-50.1 \\ \hline
\end{tabular}
\end{table*}
\endgroup

Trends for both the number of variables and clauses vary based on the type of attack and the complexity of the circuit. In the last two columns of the Table~\ref{tab:attack_res}, we list the number of correct key bits found by the cyclic SAT and BBO. We also run these attacks multiple times to see if the bits are stable, but it turns into a key guess. Notice that BBO is oracle-less, which does not utilize a SAT solver. Both Icy and BBO have a \emph{success rate of almost 50\%}, which provides a key with the correct and incorrect bits. 

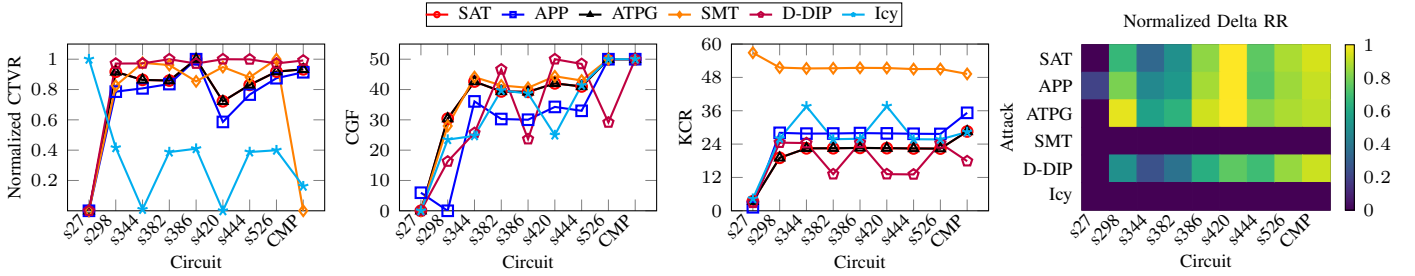
\begin{figure*}[htb]
  \centering
  \begin{adjustbox}{scale=1.0}
    \begin{tikzpicture}
      \begin{groupplot}[
        group style={group size=4 by 1, horizontal sep=1.1cm},
        width=5.0cm,
        height=3.8cm,
        xlabel={Circuit},
        ylabel style={yshift=-3pt},
        xlabel style={yshift=6pt},
        xtick={1,2,3,4,5,6,7,8,9},
        xticklabels={s27, s298, s344, s382, s386, s420, s444, s526, CMP},
        xticklabel style={rotate=45, anchor=east, font=\scriptsize},
        tick label style={font=\scriptsize},
        label style={font=\scriptsize},
        grid=none,
        title style={font=\scriptsize, yshift=-1mm},
        legend style={font=\scriptsize, at={(2.32,1.25)}, anchor=north,
                      legend columns=6, inner sep=-0.7pt, row sep=-2pt,
                      column sep=1.5pt, nodes={scale=1.0, transform shape},
                      legend image post style={scale=0.5}}
      ]

    \nextgroupplot[ylabel={Normalized CTVR}, ylabel style={yshift=0pt}, ymin=0.0, ymax=1.1, ytick={0.2, 0.4, 0.6, 0.8, 1.0}]
    \addplot[mark=o, thick, red] coordinates {(1,0.0000) (2,0.9172) (3,0.8636) (4,0.8591) (5,1.0000) (6,0.7222) (7,0.8305) (8,0.9179) (9,0.9332)};
    \addplot[mark=square, thick, blue] coordinates {(1,0.0000) (2,0.7860) (3,0.8077) (4,0.8360) (5,1.0000) (6,0.5862) (7,0.7667) (8,0.8747) (9,0.9138)};
    \addplot[mark=triangle, thick, black] coordinates {(1,0.0000) (2,0.9162) (3,0.8636) (4,0.8591) (5,1.0000) (6,0.7222) (7,0.8305) (8,0.9179) (9,0.9332)};
    \addplot[mark=diamond, thick, orange] coordinates {(1,0.0000) (2,0.8249) (3,0.9760) (4,0.9614) (5,0.8538) (6,0.9483) (7,0.8806) (8,1.0000) (9,0.0000)};
    \addplot[mark=pentagon, thick, purple] coordinates {(1,0.0000) (2,0.9713) (3,0.9721) (4,1.0000) (5,0.9713) (6,0.9999) (7,0.9988) (8,0.9721) (9,0.9915)};
    \addplot[mark=star, thick, cyan] coordinates {(1,1.0000) (2,0.4160) (3,0.0090) (4,0.3872) (5,0.4097) (6,0.0000) (7,0.3877) (8,0.3995) (9,0.1620)};
    \legend{SAT, APP, ATPG, SMT, D-DIP, Icy} 
      
    \nextgroupplot[ylabel={CGF}, xshift=-3pt, ylabel style={yshift=0pt}, ymin=0.0, ymax=55.0, ytick={0.0, 10.0, 20.0, 30.0, 40.0, 50.0}]
    \addplot[mark=o, thick, red] coordinates {(1,0.0000) (2,30.3964) (3,42.6331) (4,39.3636) (5,39.2164) (6,42.0284) (7,41.0098) (8,50.0000) (9,50.0000)};
    \addplot[mark=square, thick, blue] coordinates {(1,5.9568) (2,0.0000) (3,36.0773) (4,30.2726) (5,30.0391) (6,34.2637) (7,32.9822) (8,50.0000) (9,50.0000)};
    \addplot[mark=triangle, thick, black] coordinates {(1,0.0000) (2,30.3918) (3,42.6331) (4,39.3636) (5,39.2164) (6,42.0284) (7,41.0098) (8,50.0000) (9,50.0000)};
    \addplot[mark=diamond, thick, orange] coordinates {(1,0.0000) (2,27.9792) (3,44.1002) (4,41.3445) (5,40.5703) (6,44.3644) (7,42.9731) (8,50.0000) (9,50.0000)};
    \addplot[mark=pentagon, thick, purple] coordinates {(1,0.0000) (2,16.2629) (3,25.6896) (4,46.6405) (5,23.6380) (6,50.0000) (7,48.4421) (8,29.1331) (9,50.0000)};
    \addplot[mark=star, thick, cyan] coordinates {(1,0.0000) (2,23.4739) (3,24.7379) (4,39.7081) (5,38.6811) (6,24.9726) (7,41.3986) (8,50.0000) (9,50.0000)};

    \nextgroupplot[ylabel={KCR},  xshift=-3pt, ylabel style={yshift=-0.3pt}, ymin=0.0, ymax=60.0, ytick={0.0, 12.0, 24.0, 36.0, 48.0, 60.0}]
    \addplot[mark=o, thick, red] coordinates {(1,3.4260) (2,19.0614) (3,22.4676) (4,22.4997) (5,22.6245) (6,22.5334) (7,22.4248) (8,22.3756) (9,28.5354)};
    \addplot[mark=square, thick, blue] coordinates {(1,1.2663) (2,28.0520) (3,27.7700) (4,27.8089) (5,27.9631) (6,27.8503) (7,27.7179) (8,27.6577) (9,35.2517)};
    \addplot[mark=triangle, thick, black] coordinates {(1,3.4260) (2,19.0614) (3,22.4676) (4,22.4997) (5,22.6245) (6,22.5334) (7,22.4248) (8,22.3756) (9,28.5354)};
    \addplot[mark=diamond, thick, orange] coordinates {(1,56.8765) (2,51.5362) (3,51.2179) (4,51.2923) (5,51.4425) (6,51.3715) (7,50.9990) (8,51.0263) (9,49.2486)};
    \addplot[mark=pentagon, thick, purple] coordinates {(1,2.8859) (2,24.5169) (3,24.3697) (4,13.1630) (5,24.4684) (6,13.1843) (7,13.0857) (8,24.2822) (9,17.8576)};
    \addplot[mark=star, thick, cyan] coordinates {(1,4.3201) (2,26.0183) (3,37.5668) (4,25.7460) (5,25.9299) (6,37.6427) (7,25.7490) (8,25.6906) (9,28.3335)};


    \nextgroupplot[
  title={Normalized Delta RR},
  xshift=4pt, ylabel style={yshift=0pt},
  xlabel={Circuit},
  ylabel={Attack},
  enlargelimits=false,
  scale only axis=true,
  width=3.3cm,
  height=2.2cm,
  xtick={0,...,8},
  ytick={0,...,5},
  xticklabels={s27, s298, s344, s382, s386, s420, s444, s526, CMP},
  yticklabels={SAT, APP, ATPG, SMT, D-DIP, Icy},
    xticklabel style={rotate=-45, anchor=east, font=\scriptsize},
    tick label style={font=\scriptsize},
  y dir=reverse,
  xticklabel style={rotate=45, anchor=east, font=\scriptsize},
  yticklabel style={font=\scriptsize},
  tick style={draw=none},
  point meta min=0,
  point meta max=1,
  colormap name=viridis,
  colorbar,
  colorbar style={
    yticklabel style={font=\scriptsize},
    ytick style={draw=none},
    width=0.12cm,     
    xshift=-3pt 
  },
]

\addplot [matrix plot*, mesh/cols=9, point meta=explicit] table[meta=val] {
x y val
0 0 0.000
1 0 0.666
2 0 0.333
3 0 0.500
4 0 0.833
5 0 1.000
6 0 0.750
7 0 0.875
8 0 0.938

0 1 0.200
1 1 0.800
2 1 0.467
3 1 0.533
4 1 0.867
5 1 1.000
6 1 0.733
7 1 0.855
8 1 0.916

0 2 0.000
1 2 0.964
2 2 0.571
3 2 0.643
4 2 0.929
5 2 1.000
6 2 0.823
7 2 0.877
8 2 0.904

0 3 0.000
1 3 0.000
2 3 0.000
3 3 0.000
4 3 0.000
5 3 0.000
6 3 0.000
7 3 0.000
8 3 0.000

0 4 0.000
1 4 0.500
2 4 0.250
3 4 0.375
4 4 0.625
5 4 0.750
6 4 0.687
7 4 0.844
8 4 0.923

0 5 0.000
1 5 0.000
2 5 0.000
3 5 0.000
4 5 0.000
5 5 0.000
6 5 0.000
7 5 0.000
8 5 0.000
};
      \end{groupplot}
    \end{tikzpicture}
  \end{adjustbox}
  \caption{Attack-trace metrics CTVR, CGF, KCR, and Redundancy Ratio (RR) across ISCAS'89 benchmarks.}
  \vspace{-9pt}
  \label{fig:ctvr_cgf_kcr_rr}
\end{figure*}

Next, we leverage the results from Table~\ref{tab:attack_res} and summarize the attack metrics in Fig.~\ref{fig:ctvr_cgf_kcr_rr}. The first three panels depict the normalized CTVR, CGF, and KCR. Normalized CTVR reflects the density of SAT clauses relative to the number of variables -- a consistent trend above 0.8 in most circuits suggests \emph{constraint-dense} CNF instances in our evaluation flow with the exception of \texttt{s27}\footnote{Fixed scripts were used to prepare the CNF for each attack with identical settings (solver version, random seeds, time etc.), and to collect outputs.}. The Icy attack shows significantly lower values due to its disruptive effect on cyclic logic structures. The second panel, showing CGF, reveals the structural overhead added by obfuscation. Larger circuits demonstrate steeper CGF growth, especially under ATPG and SMT attacks. Interestingly, D-DIP maintains lower CGF for larger designs, suggesting more efficient clause management in obfuscated CNFs. 

In the third panel, KCR provides context on how the key size compares to the SAT variable count for each circuit. The SMT attack shows a higher average value whereas other attacks appear to oscillate near the 25 value. The fourth panel presents a heatmap capturing how redundancy varies across attack types for each circuit. This metric highlights how structural redundancy interacts differently with distinct attack strategies. Warmer colors represent greater variation in RR across attack variants for each circuit, whereas cooler colors denote more consistent redundancy behavior across attacks.


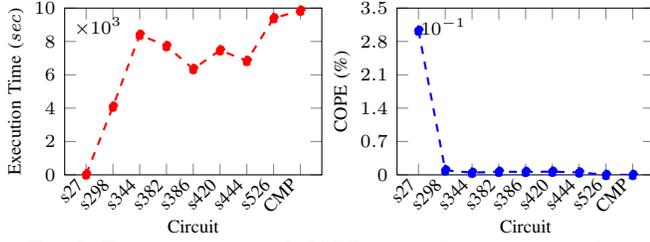
\begin{figure}[tb]
  \centering
  \begin{adjustbox}{scale=1.0}
    \begin{tikzpicture}
      \begin{groupplot}[
        group style={group size=2 by 1, horizontal sep=1cm},
        width=5.0cm,
        height=3.8cm,
        xlabel={Circuit},
        ylabel style={yshift=-3pt},
        xlabel style={yshift=6pt},
        xtick={1,2,3,4,5,6,7,8,9},
        xticklabels={s27, s298, s344, s382, s386, s420, s444, s526, CMP},
        xticklabel style={rotate=45, anchor=east, font=\scriptsize},
        tick label style={font=\scriptsize},
        label style={font=\scriptsize},
        title style={font=\scriptsize, yshift=-1mm}
      ]

      \nextgroupplot[ylabel={Execution Time ($\it{sec}$)}, ytick={0, 2000, 4000, 6000, 8000, 10000}, ymin=0, ymax=10000, yticklabel=\pgfmathparse{\tick/1000}\pgfmathprintnumber{\pgfmathresult}, scaled ticks=false]
      \addplot[mark=*, mark options={scale=0.8}, dashed, thick, red] coordinates {
        (1, 28.62) (2, 4090.29) (3, 8402.33) (4, 7726.20) (5, 6349.12) (6, 7474.90) (7, 6838.99) (8, 9397.75) (9, 9850.00)
      };
      \node[anchor=north west, font=\scriptsize] at (rel axis cs:0,1) {$\times 10^{3}$};

      \nextgroupplot[ylabel={COPE (\%)}, ytick={0.00, 0.07, 0.14, 0.21, 0.28, 0.35}, ymin=0.0, ymax=0.35, yticklabel=\pgfmathparse{\tick*10}\pgfmathprintnumber{\pgfmathresult}, scaled ticks=false]
      \addplot[mark=*, mark options={scale=0.8}, dashed, thick, blue] coordinates {
        (1, 0.3029) (2, 0.0096) (3, 0.0046) (4, 0.0066) (5, 0.0063) (6, 0.0067) (7, 0.0051) (8, 0.0000) (9, 0.0000)
      };
      \node[anchor=north west, font=\scriptsize, yshift=0pt, xshift=3pt] at (rel axis cs:0,1) {$\cdot 10^{-1}$};

      \end{groupplot}
    \end{tikzpicture}
  \end{adjustbox}
  \caption{Execution time and COPE metric for various attacks.}
  \vspace{-10pt}
  \label{fig:exec_cope}
\end{figure}

In an OL scenario, we evaluate the SCOPE attack~\cite{alaql2021}, a resynthesis-based method that exploits structural vulnerabilities. Fig.~\ref{fig:exec_cope} shows runtime (left) and the COPE score (right). The execution time clearly scales with circuit complexity and key size. COPE drops sharply after \emph{s27} and then saturates near zero, indicating minimal structural leakage. 

Overall, the results highlight an important difference between theoretical security and physical realizability. UCBO prioritizes high security: universalizing the full design obfuscates structure and function, resulting in near-zero COPE and approximately 50\% key guess. This exceeds the structural indistinguishability of gate-, LUT-, and eFPGA-based redaction. However, this degree of obfuscation comes at a significant cost in PPA.

\vspace{-2pt}
\section{Conclusions}
\vspace{-1.5pt}
\textsc{UCbo} is able to resist state-of-the-art attacks, including SAT and its variants. Results considering our attack-trace metrics (CTVR$>$0.7 and CGF$>$30), demonstrate scalability and track vulnerability trends for OG attacks. Additionally, the COPE metrics, which are below 0.1\%, indicate minimal leakage in the case of OL attacks. The oracle-less attacks (BBO) and SCOPE show no adversarial advantage, while cyclic SAT (Icy) and black-box methods are limited to a maximum success rate of 50\%, i.e., random guessing. These results combined indicate that \textsc{UCbo} provides strong resistance against known attacks.

\bibliographystyle{IEEEtran}
\bibliography{obfuscation}

@IEEEtranBSTCTL{IEEEexample:BSTcontrol,
  CTLuse_forced_etal       = "yes",
  CTLmax_names_forced_etal = "3",
  CTLnames_show_etal       = "1",
    dashed = "no"
}

@phdthesis{thesis,
author={Garg,Deepali},
year={2024},
title={Secure-by-Design: A Comprehensive Approach Towards Trusted ICs},
journal={ProQuest Dissertations and Theses},
pages={106},
isbn={9798384450955},
language={English}
}

@INPROCEEDINGS{roy2008,
  author={Roy, Jarrod A. and Koushanfar, Farinaz and Markov, Igor L.},
  booktitle={2008 DATE}, 
  title={EPIC: Ending Piracy of Integrated Circuits}, 
  year={2008},
  volume={},
  number={},
  pages={1069-1074},
  doi={10.1109/DATE.2008.4484823}
}

@ARTICLE{alaql2021,
  author={Alaql, Abdulrahman and Rahman, Md Moshiur and Bhunia, Swarup},
  journal={IEEE TVLSI},
  title={\uppercase{SCOPE}: Synthesis-Based Constant Propagation Attack on Logic Locking}, 
  year={2021},
  volume={29},
  number={8},
  pages={1529-1542},
  doi={10.1109/TVLSI.2021.3089555}}

@INPROCEEDINGS{subramanyan2015,
  author={Subramanyan, Pramod and Ray, Sayak and Malik, Sharad},
  booktitle={2015 IEEE HOST},
  title={Evaluating the security of logic encryption algorithms}, 
  year={2015},
  volume={},
  number={},
  pages={137-143},
  doi={10.1109/HST.2015.7140252}}

@INPROCEEDINGS{rajendran2012,
  author={Rajendran, Jeyavijayan and Pino, Youngok and Sinanoglu, Ozgur and Karri, Ramesh},
  booktitle={2012 IEEE/ACM DAC}, 
  title={Security analysis of logic obfuscation}, 
  year={2012},
  volume={},
  number={},
  pages={83-89},
  doi={10.1145/2228360.2228377}}

@ARTICLE{chian2020,
  author={Chiang, Hsiao-Yu and Chen, Yung-Chih and Ji, De-Xuan and Yang, Xiang-Min and Lin, Chia-Chun and Wang, Chun-Yao},
  journal={IEEE TCAD}, 
  title={LOOPLock: Logic Optimization-Based Cyclic Logic Locking}, 
  year={2020},
  volume={39},
  number={10},
  pages={2178-2191},
  doi={10.1109/TCAD.2019.2960351}}

@ARTICLE{abideen2024,
author={Abideen, Zain Ul and Gokulanathan, Sumathi and Aljafar, M. J. and Pagliarini, Samuel},
journal={ACM Computing Surveys}, 
title={An Overview of \uppercase{FPGA}-inspired Obfuscation Techniques}, 
year={2024},
volume={56},
number={12},
pages={1-35},
doi={10.1145/3677118}
}

@ARTICLE{ramesh2010,
  author={Karri, Ramesh and Rajendran, Jeyavijayan and Rosenfeld, Kurt and Tehranipoor, Mohammad},
  journal={Computer}, 
  title={Trustworthy Hardware: Identifying and Classifying Hardware Trojans}, 
  year={2010},
  volume={43},
  number={10},
  pages={39-46},
  doi={10.1109/MC.2010.299}}

@inproceedings{alkabani2007,
author = {Alkabani, Yousra M. and Koushanfar, Farinaz},
title = {Active Hardware Metering for Intellectual Property Protection and Security},
year = {2007},
address = {USA},
booktitle = {16th USENIX Security Symposium},
articleno = {20},
numpages = {16},
location = {Boston, MA},
series = {SS'07}
}

@inproceedings{shen2017,
author = {Shen, Yuanqi and Zhou, Hai},
title = {Double DIP: Re-Evaluating Security of Logic Encryption Algorithms},
year = {2017},
pages = {179–184},
numpages = {6},
location = {Banff, Alberta, Canada},
series = {GLSVLSI '17}
}

@INPROCEEDINGS{rezaei2022,
  author={Rezaei, Amin and Afsharmazayejani, Raheel and Maynard, Jordan},
  booktitle={2022 IEEE/ACM ICCAD}, 
  title={Evaluating the Security of eFPGA-based Redaction Algorithms}, 
  year={2022},
  volume={},
  number={},
  pages={1-7},
  doi={}}

@inproceedings {zhaokun2023,
author = {Zhaokun Han and Mohammed Shayan and Aneesh Dixit and Mustafa Shihab and Yiorgos Makris and Jeyavijayan (JV) Rajendran},
title = {{FuncTeller}: How Well Does {eFPGA} Hide Functionality?},
booktitle = {32nd USENIX Security Symposium},
year = {2023},
address = {Anaheim, CA},
pages = {5809--5826}
}

@inproceedings{valiant1976,
author = {Valiant, Leslie G.},
title = {Universal circuits (Preliminary Report)},
year = {1976},
booktitle = {Proceedings of the Eighth Annual ACM Symposium on Theory of Computing},
pages = {196–203},
numpages = {8}
}

@misc{elisaweta2022,
      author = {Elisaweta Masserova and Deepali Garg and Ken Mai and Lawrence Pileggi and Vipul Goyal and Bryan Parno},
      title = {Logic Locking - Connecting Theory and Practice},
      howpublished = {Cryptology {ePrint} Archive},
      year = {2022}
}

@inproceedings{disser2023,
author = {Disser, Yann and G\"{u}nther, Daniel and Schneider, Thomas and Stillger, Maximilian and Wigandt, Arthur and Yalame, Hossein},
title = {Breaking the Size Barrier: Universal Circuits Meet Lookup Tables},
year = {2023},
booktitle = {ASIACRYPT},
pages = {3–37},
numpages = {35}
}

@InProceedings{crescenzo2020,
author="Di Crescenzo, Giovanni
and Sengupta, Abhrajit
and Sinanoglu, Ozgur
and Yasin, Muhammad",
title="Logic Locking of Boolean Circuits: Provable Hardware-Based Obfuscation from a Tamper-Proof Memory",
booktitle="SecITC",
year="2020",
pages="172--192"
}

@INPROCEEDINGS{shamsi2017,  author={Shamsi, Kaveh and Li, Meng and Meade, Travis and Zhao, Zheng and Pan, David Z. and Jin, Yier},  booktitle={2017 IEEE HOST},   title={AppSAT: Approximately deobfuscating integrated circuits},   year={2017},  volume={},  number={},  pages={95-100},  doi={10.1109/HST.2017.7951805}}

@INPROCEEDINGS{Abideen2025b,
  author={Ul Abideen, Zain and Aksoy, Levent and Pagliarini, Samuel},
  booktitle={2025 DATE}, 
  title={Late Breaking Results: Is Reconfigurable-Based Obfuscation Secure?}, 
  year={2025},
  volume={},
  number={},
  pages={1-2},
  doi={10.23919/DATE64628.2025.10993007}}

@article{zamiri2018, title={SMT Attack: Next Generation Attack on Obfuscated Circuits with Capabilities and Performance Beyond the SAT Attacks}, volume={2019},  DOI={10.13154/tches.v2019.i1.97-122}, number={1}, journal={IACR TCHES}, 
author={Azar, Kimia Zamiri and Kamali, Hadi Mardani and Homayoun, Houman and Sasan, Avesta}, 
year={2018}, month={Nov.}, pages={97–122} }

@INPROCEEDINGS{shamsi2019b,
  author={Shamsi, Kaveh and Pan, David Z. and Jin, Yier},
  booktitle={2019 IEEE HOST}, 
  title={On the Impossibility of Approximation-Resilient Circuit Locking}, 
  year={2019},
  volume={},
  number={},
  pages={161-170},
  doi={10.1109/HST.2019.8741035}}
\end{document}